\documentclass[journal]{IEEEtran}
\pdfoutput=1
\usepackage{amsmath,amsfonts}
\usepackage{algorithmic}
\usepackage{algorithm}
\usepackage{array}
\usepackage[caption=false,font=normalsize,labelfont=sf,textfont=sf]{subfig}
\usepackage{textcomp}
\usepackage{stfloats}
\usepackage{url}
\usepackage[hidelinks]{hyperref}
\usepackage{verbatim}
\usepackage{graphicx}
\usepackage{cite}
\usepackage{glossaries}
\usepackage{siunitx}
\usepackage{flushend}
\usepackage{listings}
\usepackage{xcolor}
\usepackage{textcomp}

\definecolor{codegreen}{rgb}{0.25,0.47,0.47}
\definecolor{codegray}{rgb}{0.5,0.5,0.5}
\definecolor{codepurple}{rgb}{0.71,0.21,0.19}
\definecolor{backcolor}{rgb}{0.95,0.95,0.92}
\definecolor{emphcolor}{rgb}{0.08, 0.29, 0.62}

\lstdefinestyle{mystyle}{
    backgroundcolor=\color{white},   
    commentstyle=\color{codegreen},
    keywordstyle=\color{magenta},
    numberstyle=\tiny\color{codegray},
    stringstyle=\color{codepurple},
    basicstyle=\ttfamily\footnotesize,
    breakatwhitespace=false,         
    breaklines=true,                 
    captionpos=b,                    
    keepspaces=true,                 
    numbers=left,                    
    numbersep=5pt,                  
    showspaces=false,                
    showstringspaces=false,
    showtabs=false,                  
    tabsize=2,
    frame = single,
    emph={
        Constellation,
        BinarySource,
        LDPC5GEncoder,
        LDPC5GDecoder,
        Mapper,
        AWGN,
        Demapper,
        compute_ber,
        NeuralDemapper,
        BinaryCrossentropy,
        Dense,
        Layer,
        load_scene,
        Transmitter,
        Receiver,
        Paths2CIR
        },
    emphstyle=\color{emphcolor}
}
\newacronym{ACM}{ACM}{adaptive coding and modulation}
\newacronym{ADC}{ADC}{analog-to-digital conversion}
\newacronym{AGC}{AGC}{automatic gain control}
\newacronym{AOA}{AOA}{angle of arrival}
\newacronym{API}{API}{application programming interface}
\newacronym{AWGN}{AWGN}{additive white Gaussian noise}
\newacronym{AI}{AI}{artificial intelligence}
\newacronym{BER}{BER}{bit error rate}
\newacronym{BEC}{BEC}{binary erasure channel}
\newacronym{BLER}{BLER}{block error rate}
\newacronym{BP}{BP}{backpropagation}
\newacronym{BPA}{BP}{belief propagation}
\newacronym{BPTT}{BPTT}{back-propagation through time}
\newacronym{CFO}{CFO}{carrier frequency offset}
\newacronym{CN}{CN}{check node}
\newacronym{CNN}{CNN}{convolutional neural network}
\newacronym{CND}{CND}{check node decoder}
\newacronym{ConvNet}{ConvNet}{convolutional neural network}
\newacronym{CP}{CP}{cyclic prefix}
\newacronym{CRC}{CRC}{cyclic redundancy check}
\newacronym{CSI}{CSI}{channel state information}
\newacronym{DAC}{DAC}{digital-to-analog conversion}
\newacronym{DL}{DL}{deep learning}
\newacronym{DFT}{DFT}{discrete Fourier transform}
\newacronym{DOA}{DOA}{direction of arrival}
\newacronym{FFT}{FFT}{fast Fourier transform}
\newacronym{FEC}{FEC}{forward error correction}
\newacronym{GAN}{GAN}{generative adversarial network}
\newacronym{GRU}{GRU}{gated recurrent unit}
\newacronym{GPU}{GPU}{graphic processing unit}
\newacronym{GPS}{GPS}{global positioning system}
\newacronym{CPU}{CPU}{central processing unit}
\newacronym{HDPC}{HDPC}{high-density parity-check}
\newacronym{iid}{i.i.d.\@}{independent and identically distributed}
\newacronym{IFFT}{IFFT}{inverse fast Fourier transform}
\newacronym{KL}{KL}{Kullback-Leibler}
\newacronym{LDPC}{LDPC}{low-density parity-check}
\newacronym{LLR}{LLR}{log-likelihood ratio}
\newacronym{LOS}{LoS}{line of sight}
\newacronym{LS}{LS}{least squares}
\newacronym{LSTM}{LSTM}{long short-term memory}
\newacronym{LMMSE}{LMMSE}{linear minimum mean squared error}
\newacronym{MIMO}{MIMO}{multiple-input multiple-output}
\newacronym{ML}{ML}{machine learning}
\newacronym{MLE}{MLE}{maximum likelihood estimation}
\newacronym{MLP}{MLP}{multilayer perceptron}
\newacronym{MRC}{MRC}{maximum ratio combining}
\newacronym{MSE}{MSE}{mean squared error}
\newacronym{MPA}{MPA}{message passing algorithm}
\newacronym{MMSE}{MMSE}{minimum mean squared error}
\newacronym{NLP}{NLP}{natural language processing}
\newacronym{NN}{NN}{neural network}
\newacronym{NLOS}{NLoS}{non-line of sight}
\newacronym{OFDM}{OFDM}{orthogonal frequency-division multiplexing}
\newacronym{OSS}{OSS}{open-source software}
\newacronym{pdf}{pdf}{probability density function}
\newacronym{pmf}{pmf}{probability mass function}
\newacronym{ReLU}{ReLU}{rectified linear unit}
\newglossaryentry{RIS}
{
  name={RIS},
  description={reconfigurable intelligent surface},
  first={\glsentrydesc{RIS} (\glsentrytext{RIS})},
  firstplural={\glsentrydesc{RIS}s (\glsentrytext{RIS})},
  plural={RIS}
}
\newacronym{RL}{RL}{reinforcement learning}
\newacronym{RNN}{RNN}{recurrent neural network}
\newacronym{RTN}{RTN}{radio transformer network}
\newacronym{SDR}{SDR}{software defined radio}
\newacronym{SER}{SER}{symbol error rate}
\newacronym{SFO}{SFO}{sampling frequency offset}
\newacronym{SNR}{SNR}{signal-to-noise ratio}
\newacronym{SINR}{SINR}{signal-to-interference-plus-noise ratio}
\newacronym{SGD}{SGD}{stochastic gradient descent}
\newacronym{SPA}{SPA}{sum product algorithm}
\newacronym{VN}{VN}{variable node}
\newacronym{VND}{VND}{variable node decoder}
\newacronym{SVM}{SVM}{support vector machine}
\newacronym{TDOA}{TDOA}{time difference of arrival}
\newacronym{TOA}{TOA}{time of arrival}
\newacronym{UAV}{UAV}{unmanned aerial vehicle}
\newacronym{wrt}{w.r.t.\@}{with respect to}
\newacronym{ZF}{ZF}{zero forcing}

\begin{document}

\title{Sionna: An Open-Source Library for Next-Generation Physical Layer Research}

\author{
    Jakob Hoydis,    
    Sebastian Cammerer,
    Fay\c{c}al A\"it Aoudia,
    Avinash Vem,\\
    Nikolaus Binder,
    Guillermo Marcus,
    Alexander Keller
}
\maketitle

\begin{abstract}
    Sionna\texttrademark{} is a GPU-accelerated open-source library for link-level simulations based on TensorFlow. It enables the rapid prototyping of complex communication system architectures and provides native support for the integration of neural networks. Sionna implements a wide breadth of carefully tested state-of-the-art algorithms that can be used for benchmarking and end-to-end performance evaluation. This allows researchers to focus on their research, making it more impactful and reproducible, while saving time implementing components outside their area of expertise. This white paper provides a brief introduction to Sionna, explains its design principles and features, as well as future extensions, such as integrated ray tracing and custom CUDA kernels. We believe that Sionna is a valuable tool for research on next-generation communication systems, such as 6G, and we welcome contributions from our community.
  \end{abstract}
  \glsresetall

\section{Introduction}
\IEEEPARstart{W}{hile} 5G is deployed around the globe and the work areas for its future evolution called 5G Advanced (Release 18) have just been approved \cite{rel18-a}, researchers in academia and industry have already started to define visions and key technologies for 6G, e.g., \cite{hexa-x}. One recurring vision in many papers is that of creating digital twin worlds to connect physical and biological entities and to enable new mixed-reality experiences such as the Metaverse \cite{6g}.

Some of the most prominent 6G research topics are communication in the Terahertz band \cite{thz}, 
cell-free, holographic, and very large aperture \gls{MIMO} \cite{mimo}, \gls{UAV} and satellite communication \cite{uav}, \gls{ML}-based air interface design \cite{aiai}, semantic communication \cite{semantic}, \glspl{RIS} \cite{ris}, sensing and localization \cite{sensing}, digital twins \cite{dtn}, computer vision for wireless \cite{wivi}, as well as federated learning over wireless \cite{fl}.
Interestingly, research on most of the above topics requires one or more of the following:
\begin{itemize}
    \item \textbf{Data from specific radio environments:}
    While the widely used stochastic channel models, e.g., \cite{38901}, are well suited to capture the behavior of a certain class of environment, they cannot be used for any application that requires the simulation of a \emph{specific} environment, e.g., the optimal configuration of a particular \gls{RIS}. Radar-based sensing and localization as well as computer vision-aided applications only work if there is a spatially consistent correspondence between physical location and channel impulse response (or visual input). This can only be achieved through either ray tracing or extensive measurement campaigns.

    \item \textbf{Native integration of \gls{ML}:}
    \gls{ML} and, particularly, \glspl{NN} are expected to play an increasingly important role for the implementation of transceiver algorithms and maybe even the 6G air interface design \cite{rel18-a, aiai, 6gflagshipML}. Research on such topics does not only benefit tremendously from a tight integration of \gls{ML}, e.g., \cite{tensorflow, keras}, and link-level simulation tools, but also from automatic gradient computation through the entire system which allows for seamless integration of \glspl{NN}.

    \item \textbf{Very high detail or scale:}
    Link-level simulations of 6G systems require unprecedented modeling accuracy and scale.
    While cell-free or very large aperture \gls{MIMO} systems need the computation of a very large number of signal propagation paths, Terahertz communication systems have complex channel models accounting for nonlinear hardware impairments. Moreover, the full potential of \gls{ML}-enhanced algorithms will only be unleashed with very realistic simulations that account for detail that has been ignored in the past due to prohibitive simulation complexity and for mathematical tractability.
\end{itemize}

Due to the reasons above, we firmly believe that the breakthroughs required for 6G can only be achieved if our community has access to new tools for physical layer research. That is why we have started to develop a new and open-source link-level simulator called Sionna.\footnote{\url{https://github.com/NVlabs/sionna}}

\section{Motivation \& Background}

\subsection{Why we decided to create Sionna}
Sionna\footnote{Sionna is a goddess in Irish mythology and the namesake of the River Shannon, see \url{https://en.wikipedia.org/wiki/Shannon_(given_name)}.} is written by researchers for researchers in communications and aims at making our work more efficient:
\begin{itemize}
    \item \textbf{Rapid prototyping:}
        Sionna provides a high-level Python \gls{API} to rapidly model complex communication systems from end-to-end while allowing one to easily adapt the parts a new research idea is about. \Gls{GPU} acceleration makes it super fast and enables interactive exploration, e.g., in Jupyter notebooks \cite{jupyter}.

    \item \textbf{Benchmarking against the state of the art:}
        Sionna provides many carefully tested standard processing blocks and state-of-the-art algorithms that can be used for performance benchmarking. This reduces the time spent implementing auxiliary components that one's research is not about.

    \item \textbf{Realistic industry-grade evaluations:}
        Experts in one domain, say channel estimation, do not necessarily have the time, tools, or background to evaluate their algorithms for end-to-end performance, e.g, coded \gls{BLER} of a 5G Polar code over a realistic  3GPP channel model. With Sionna, one can make such evaluations with a few additional lines of code. If needed, simulations can be scaled across large multi-\gls{GPU} setups.

    \item \textbf{Native support of \glspl{NN}:}
        Sionna enables seamless integration of \glspl{NN} in the physical layer signal processing chain. As most building blocks are differentiable, gradients can be backpropagated through an entire system, which is the key enabler for end-to-end learning of new \gls{AI}-defined air interfaces. Sionna also eliminates the need for different tools for data generation, training, and evaluation.

    \item \textbf{Addressing 6G research needs:}
        Sionna will allow for the use of ray tracing, datasets, and generative models instead of stochastic channel models to enable research on many novel topics, such as joint communication and sensing, vision-aided wireless communications, intelligent reflecting surfaces, semantic communications, and digital twins. Also, a THz channel model is under development. Community contributions to components are welcome.
    
    \item \textbf{Reproducibility:}
        With Sionna, it is straight-forward to make research reproducible by others. We encourage all users to publish their Sionna-based code along with their research papers and contribute novel components so that they can be reused by others. This will also make it easier to compare algorithms under the same conditions.

\end{itemize}

\subsection{Related \gls{OSS}}
Commercial software products, such as Matlab \cite{matlab}, play an important role for simulation-based research in our field. There is also a growing body of well-maintained \gls{OSS} projects serving different purposes, such as channel simulation \cite{quadriga, nyusim}, link-level simulations \cite{Vienna5GLLS, aff3ct, hermespy}, system-level simulations \cite{Vienna5GSLS}, network simulations \cite{ns-3, omnet++}, and hardware experimentation \cite{gnuradio, oai}. There are also several open experimental platforms, such as the Platforms for Advanced Wireless Research (PAWR) in the US \cite{pawr} and the OneLab Platforms \cite{onelab} in Europe. These allow for reproducible large-scale experiments in controlled and/or realistic environments. Not all the mentioned \gls{OSS}, e.g., \cite{quadriga, nyusim, hermespy}, accept contributions from external parties though.

Sionna is a tool for link-level simulations with in-built channel simulator. It has hence some similarities with Quadriga \cite{quadriga} and NYUSIM \cite{nyusim} for the generation of channel impulse responses, as well as with the Vienna 5G Link-Level Simulator \cite{Vienna5GLLS} and HermesPy \cite{hermespy} for physical layer processing. It differs, however, through its end-to-end differentiability, native support of \glspl{NN} and \glspl{GPU}, as well as ray tracing capabilities. Sionna can be scaled to large multi-\gls{GPU} setups to simulate, e.g., realistic multi-cell Massive MIMO systems.

\section{A Sionna Primer}
Sionna is written in Python using TensorFlow \cite{tensorflow} and the Keras \gls{API} \cite{keras}. Exceptions are the custom CUDA kernels and ray tracing capabilities that will be described in Sec.~\ref{sec:extensions}.

\subsection{Design principles}
One of the key differences between Sionna and alternative link-level simulators is 
that all algorithms are implemented using high-dimensional Tensors, e.g., of shape

\texttt{[batch\_size, num\_tx,..., fft\_size]}

\noindent where the first dimension is the \textit{batch size}, representing different independent Monte Carlo trials that are executed in parallel. For-loops are generally avoided.
This representation is borrowed from the field of deep learning and leads to ``embarrassingly parallel'' workloads that can be efficiently executed on \glspl{GPU}. If no \gls{GPU} is available, Sionna will also run on (multiple) \glspl{CPU}. 

Another design principle of Sionna is that all components are implemented as independent Keras layers \cite{keras}. This has the advantages that (i) complex system models can be constructed by simply connecting the desired layers, (ii) components can be easily replaced by \glspl{NN}, and (iii) gradients are automatically computed by TensorFlow \cite{tensorflow} which is a key enabler for end-to-end learning of communication systems. Both Keras' sequential and functional \gls{API}s can be used.

Sionna's default data type is \texttt{tf.complex64}, i.e., 32bit-precision for each complex dimension. For certain applications, e.g., optical communications, the precision of all or a few selected layers can be increased to \texttt{tf.complex128}. Note that this also doubles their memory requirements.

Sionna supports both of TensorFlow's execution modes: eager and graph execution. The former is very useful for interactive development of new components, while the latter provides improved memory efficiency and reduced runtime. In addition, all of Sionna's layers (apart from a few exceptions) support just-in-time compilation using XLA (accelerated linear algebra) \cite{xla} for further speed-ups. 

\subsection{Features}
The first public release of Sionna (v0.8.0) implements the following list of features:\\
\textbf{\Gls{FEC}:}
\begin{itemize}
        \item 5G LDPC codes including rate matching \cite{38212}
        \item 5G Polar codes including rate matching \cite{38212}
        \item \Gls{CRC} \cite{38212}
        \item Reed-Muller \& Convolutional codes
        \item Interleaving \& Scrambling
        \item Belief propagation (BP) decoder and variants
        \item SC, SCL, and SCL-CRC Polar decoders
        \item Viterbi decoder
        \item Demapper with prior
        \item EXIT chart simulations
        \item Import of partity-check matrices in \emph{alist} format
\end{itemize}

\noindent\textbf{Channel models}:
\begin{itemize}
    \item \Gls{AWGN} channel
    \item Flat-fading channel models with antenna correlation
    \item 3GPP 38.901 TDL, CDL, UMa, UMi, RMa models \cite{38901}
    \item Import of channel impulse response from datasets
    \item Channel output computed in time or frequency domain
\end{itemize}\pagebreak

\noindent\textbf{\Gls{MIMO} processing:}
\begin{itemize}
    \item Multiuser \& multicell \gls{MIMO} support
    \item 3GPP 38.901 \& custom antenna arrays/patterns
    \item \Gls{ZF} precoding
    \item \Gls{MMSE} equalization
\end{itemize}

\noindent\textbf{\Gls{OFDM}:}
\begin{itemize}
    \item \Gls{OFDM} modulation \& demodulation
    \item Cyclic prefix insertion \& removal
    \item Flexible 5G slot-like frame structure
    \item Arbitrary pilot patterns
    \item LS channel estimation \& Nearest neighbor interpolation
\end{itemize}

Despite the fact that Sionna supports 5G-compliant channel codes and channel models, it does not try by any means to be a 5G-compliant link-level simulator. It rather allows researchers to test their algorithms under realistic conditions and to compare to state-of-the-art algorithms that are widely accepted in the industry, while having a maximum degree of flexibility and freedom. The current set of features should be seen as a good starting point that one can build on to rapidly prototype and evaluate new ideas.

\subsection{Hello, World!}
The Sionna webpage provides an extensive set of tutorial notebooks (\url{https://nvlabs.github.io/sionna/tutorials.html}) that cover in detail the various ways the library can be used. In order to get a first impression of how developing with Sionna feels like, we provide and discuss a few short code examples.

\begin{lstlisting}[language=Python, float, caption={Sionna "Hello, World!" example.}, label=lst:helloworld]
batch_size = 1024
n = 1000 # codeword length
k = 500  # information bits per codeword
m = 4 # number of bits per symbol
snr = 10
c = Constellation("qam", m)
b = BinarySource()([batch_size, k])
u = LDPC5GEncoder(k, n)(b)
x = Mapper(constellation=c)(u)
y = AWGN()([x, 1/snr])
llr = Demapper("app", constellation=c)([y, 1/snr])
b_hat = LDPC5GDecoder(LDPC5GEncoder(k, n))(llr)
compute_ber(b, b_hat)
\end{lstlisting}

Listing~\ref{lst:helloworld} shows a Sionna ``Hello, World!'' example in which the transmission of \texttt{batch\_size} LDPC codewords over an AWGN channel using 16QAM modulation is simulated. This example shows how Sionna layers are instantiated and then immediately applied to a previously defined tensor. For example, in line 8, the \texttt{LDPC5GEncoder} is instantiated with the parameters \texttt{n} and \texttt{k} and then applied to the tensor \texttt{b} which contains randomly generated information bits.
This coding style follows the functional \gls{API} of Keras.

In Listing~\ref{lst:trainable}, we have made the \texttt{Constellation} \emph{trainable} and replaced the \texttt{Demapper} by a \texttt{NeuralDemapper} which is defined in Listing~\ref{lst:neural_demapper}. What happens under the hood is that the tensor defining the constellation points has become a trainable TensorFlow Variable. It can now be tracked together with the weights of the \gls{NN} of the \texttt{NeuralDemapper} by TensorFlow's automatic differentiation feature, see \cite{autograd}. One could then define an adequate \texttt{loss} function (line 7), such as the total binary cross-entropy, and compute the gradient of this loss with respect to all trainable variables. This gradient may then be used to optimize jointly the constellation points and the \texttt{NeuralDemapper} through stochastic gradient descent. This concept is sometimes referred to as \emph{end-to-end learning} \cite{e2e} and can be applied to very complex system models.

Listing~\ref{lst:neural_demapper} shows the definition of a \texttt{NeuralDemapper} as a Keras layer which was used in Listing~\ref{lst:trainable} to replace a traditional \texttt{Demapper}. This simple combination of \gls{NN} components and traditional algorithms together with automatic gradient computation is one of the key advantages of Sionna for \gls{ML}-based research. 

\begin{lstlisting}[language=Python, float, caption={Components can be easily replaced by \glspl{NN} or made trainable.}, label=lst:trainable]
c = Constellation("qam", m, trainable=True)
b = BinarySource()([batch_size, k])
u = LDPC5GEncoder(k, n)(b)
x = Mapper(constellation=c)(u)
y = AWGN()([x, 1/snr])
llr = NeuralDemapper()([y, 1/snr])
loss = BinaryCrossentropy(from_logits=True)(u, llr)
\end{lstlisting}

\begin{lstlisting}[language=Python, float, caption={Definition of a simple neural demapper in Keras.}, label=lst:neural_demapper]
class NeuralDemapper(Layer):   
    def build(self, input_shape):
        # Initialize the neural network layers
        self._dense1 = Dense(16, activation="relu")
        self._dense2 = Dense(m)
        
    def call(self, inputs):
        y, no = inputs
        
        # Stack noise variance, real and imaginary
        # parts of each symbol. The input to the
        # neural network is [Re(y_i), Im(y_i), no].
        no = no*tf.ones(tf.shape(y)) 
        llr = tf.stack([tf.math.real(y),
                        tf.math.imag(y),
                        no], axis=-1)
        
        # Compute neural network output
        llr = self._dense1(llr)
        llr = self._dense2(llr)
        
        # Reshape to [batch_size, n]
        llr = tf.reshape(llr, [batch_size, -1])
        return llr
\end{lstlisting}

\subsection{Limitations}
As Sionna parallelizes simulations over batches, the available \gls{GPU} memory can quickly become the bottleneck. This can be circumvented by either reducing the batch size or distributing the model or batch across multiple \glspl{GPU}.

Algorithms with complex conditional logic, e.g., where examples in a batch can be treated differently,  do not lend themselves to be written efficiently using TensorFlow's Python \gls{API}. In this case, one can resort to implementing custom TensorFlow operations in C++ and CUDA (see Sec.~\ref{subsec:cuda}). At the cost of execution speed, it is also possible to write such algorithms in native Python code and wrap them as a TensorFlow operations. 

\subsection{Contributing}
Sionna is released under the Apache 2.0 license. We have chosen this license so that users do not need to worry about possible patent infringements and litigation. We accept and welcome contributions from external parties through \emph{pull request} on the Sionna GitHub repository. The only requirement is that all contributors ``sign-off'' their commits with the Developer Certificate of Origin (DCO) \cite{dco}.

\section{Future extensions}\label{sec:extensions}

\subsection{Custom CUDA kernels}\label{subsec:cuda}

We have chosen Python and TensorFlow for writing most of Sionna due to their popularity and simplicity.
However, not all signal processing algorithms and simulation routines can be easily expressed using tensors and compositions of existing TensorFlow operations. This is generally the case, whenever an algorithm might execute differently for each example in a batch, e.g., due to \texttt{if-else} statements, or requires complex indexing.
Examples comprise the Polar SCL and min-sum BP decoder, components of the 3GPP channel models, as well as the convolution with a time-varying channel impulse response.
Moreover, some algorithms expressed as the composition of TensorFlow operations might not lead to acceptable performance or memory requirements.

As an alternative, one can implement such algorithms without the constraints of the TensorFlow \gls{API} as \emph{custom} TensorFlow operations~\cite{tf-custom} using C++ and XLA~\cite{xla} or CUDA~\cite{cuda} to benefit from GPU acceleration.
We found the resulting code to be significantly more convenient, better readable, and less prone to errors.
The source code of a custom operation is compiled into a binary \emph{shared object} that can be loaded by TensorFlow from the Python code.
The custom operation can then be called from the Python TensorFlow \gls{API} in the same way as any other operation.
Although tensors are used as inputs and outputs to custom operations, any type of data structure can be used in the internal C++ implementation with CUDA.
Moreover, the CUDA-X libraries~\cite{cuda-x} can be leveraged to further benefit from efficient GPU-accelerated algorithms.
If required, gradients can be specified to make a custom operation differentiable.

\begin{lstlisting}[language=Python, float, caption={Sionna ray tracing example.}, label=lst:raytracing]
from sionna.rt import load_scene, Paths2CIR
# Load integrated scene
scene = load_scene(sionna.rt.scene.munich)

# Compute propagation paths
paths = scene.compute_paths()

# Compute coverage map
cm = scene.coverage_map()

# Render scene with coverage map and paths
scene.render(paths=paths, coverage_map=cm)

# Convert paths to channel impulse responses
p2c = Paths2CIR(sampling_frequency=1e6, scene=scene)
a, tau = p2c(paths.as_tuple())
\end{lstlisting}

\begin{figure}
\includegraphics[width=\linewidth]{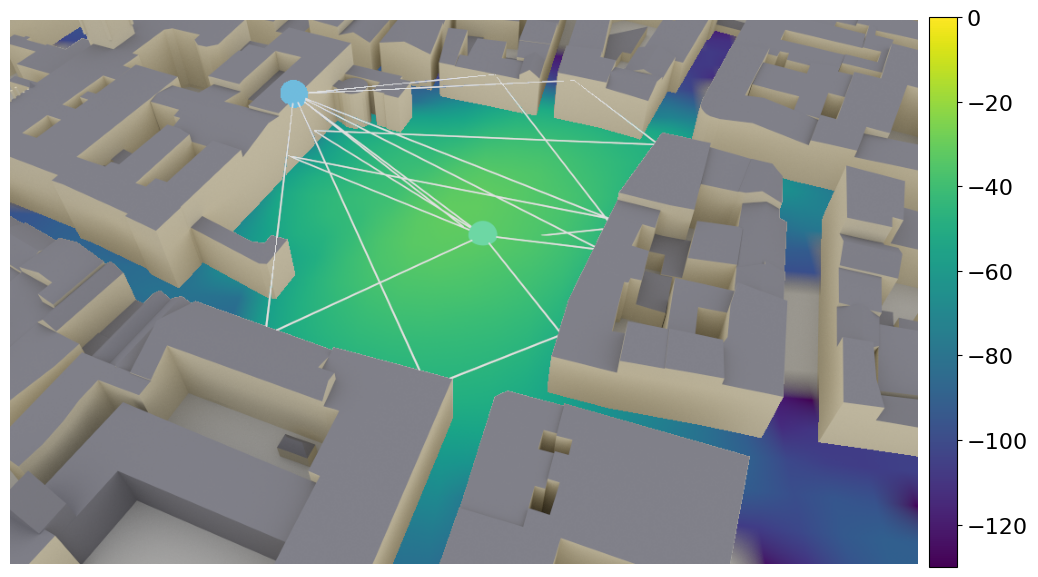}
\caption{Example of a coverage map (path gain [dB]) and ray-traced propagation paths rendered on top of a scene by Sionna RT.}
\label{Fig:RayTracing}
\end{figure}

\subsection{Ray tracing}
As mentioned earlier, many 6G topics, such as \gls{RIS} or integrated sensing and communications, require the simulation of a specific radio environment in a physically-based manner which cannot be done with stochastic channel models. For some applications, also the visual representation of a scene is required, see, e.g., \cite{viwi}.
To address these needs, Sionna RT will soon bring both of these capabilities to Sionna 
so that one can render a scene, get the desired channel impulse responses, and use 
them immediately for link-level simulations or other applications.

Listing~\ref{lst:raytracing} demonstrates the simplicity of invoking high performance ray tracing from within a Jupyter notebook. First, an integrated scene is loaded. Then, propagation paths between all transmitters and receivers are computed. Their configuration is not shown in this listing. Next, a coverage map is generated before it is rendered in the scene together with the propagation paths (see Fig.~\ref{Fig:RayTracing}). Finally, channel impulse responses between all defined transmitters and receivers are computed. These can be directly used for link-level simulations instead of a stochastic channel model. Beyond that simplicity, one can manipulate many scene parameters, such as material properties, from within Python.

\section{Conclusions}
The next decade will be marked by groundbreaking research on disruptive technologies for 6G whose exploration poses various challenges for computer simulations. 
For this reason, we have started the development of Sionna, a differentiable open-source link-level simulator with native integration of \glspl{NN} and full \gls{GPU} acceleration. Soon, Sionna will have an integrated ray tracer for scene rendering and wave propagation. We are excited about the possibilities Sionna offers and hope for wide adoption and numerous contributions from our community.

\bibliographystyle{IEEEtran}
\bibliography{IEEEabrv, bibliography}
\end{document}